\documentclass[graybox]{svmult}

\usepackage{mathptmx}
\usepackage{helvet}
\usepackage{courier}
\usepackage{type1cm}
\usepackage{makeidx}
\usepackage{graphicx}
\usepackage{multicol}
\usepackage[bottom]{footmisc}
\usepackage[numbers]{natbib}
\makeindex
\usepackage{graphicx}
\usepackage{amsmath}
\usepackage{amssymb}
\definecolor{IntLink}{rgb}{0.1,0.15,0.4}
\definecolor{ExtLink}{rgb}{0.5,0.1,0}
\usepackage[bookmarksnumbered=true,breaklinks,colorlinks,citecolor=IntLink,linkcolor=IntLink,urlcolor=ExtLink]{hyperref}

\begin{document}

\title*{Dynamical mean field theory for~oxide~heterostructures}

\author{O. Janson, Z. Zhong, G. Sangiovanni, and K. Held}
\institute{O. Janson and K. Held \at  Institute for Solid State Physics,
TU Wien, 1040 Vienna, Austria\\ \email{held@ifp.tuwien.ac.at} \and Z.
Zhong  \at Max Planck Institute for Solid State Physics,  70569
Stuttgart, Germany\\ \email{zhong@fkf.mpg.de} \and  G. Sangiovanni\at
Institut f\"ur Theoretische Physik und Astrophysik, Universit\"at
W\"urzburg, Am Hubland, D-97074 W\"urzburg, Germany }

\nopagebreak
\maketitle \abstract{Transition metal oxide heterostructures often,
but by far not always, exhibit strong electronic correlations.
State-of-the-art calculations account for these by dynamical mean
field theory  (DMFT).  We discuss the physical situations in which
DMFT is needed, not needed, and where it is actually not sufficient.
By means of an example, SrVO$_3$/SrTiO$_3$, we discuss step-by-step
and figure-by-figure a density functional theory(DFT)+DMFT
calculation. The second part reviews DFT+DMFT calculations for oxide
heterostructure focusing on titanates, nickelates, vanadates, and
ruthenates.}

\section{Introduction}
The extraordinary progress to grow heterostructures of transition metal
oxides, atomic-layer-by-atomic-layer, cherished hopes of discovering
novel physical phenomena that are non-existing in conventional
semiconductor heterostructures. Arguably the most remarkable difference
and breeding ground for these hopes is the fact that transition metal
oxides are often strongly correlated materials. Thence unconventional
and unexpected states or colossal responses are imaginable.  Hand-in-hand
with the experimental analysis of prospective correlation phenomena we
need a theoretical tool to address electronic correlations in oxide
heterostructures.  Obviously this requires going beyond density
functional theory (DFT). It  is instead the
realm of dynamical mean field theory (DMFT)
\cite{metzner89,georges92,georges96} and its merger with DFT for real
materials calculations \cite{kotliar06,held07}.

Before turning to this method, let us here in the Introduction address
the questions: For which oxide heterostructures are electronic
correlations important so that DMFT is needed? Actually, for many oxide
heterostructures  studied experimentally electronic correlations are not
that strong. In particular the LaAlO$_3$/SrTiO$_3$ prototype
\cite{ohtomo04}, which has been at the focus of the experimental efforts
at the dawn of the research field, is not strongly correlated. The
reason for this is that SrTiO$_3$ is a band insulator with empty Ti-$d$
orbitals.  Through the polar catastrophe or oxygen defects, these
Ti-bands are doped, but only slightly. With only a few charge carriers in
the Ti-$d$ orbitals we are far away from an integer filling of the $d$
orbitals, and the $d$ electrons can quite freely move around the
interface without often seeing a Ti site already occupied with a
$d$-electron and hence prone to a strong Coulomb interaction.  
Therefore not surprisingly DFT or even a simple tight binding
modeling \cite{zhong13} are sufficient to reproduce or predict the
angular resolved photoemission spectroscopy (PES, ARPES) spectra
\cite{yoshimatsu11,santander11,wang14}.  Let us note that there
might be localization of (some of) the $d$-electrons at oxygen
defects \cite{berner13prl,berner13prb,behrmann15,jeschke15}.
For this localization electronic correlations play a role.

In general, electronic correlations are strong  whenever we are at or
close to an integer filling of the $d$-orbitals in some of the layers or
at some of the sites. In this situation the Coulomb repulsion $U$ is
strong and hence important. It suppresses the mobility of the charge
carriers, leads to a strongly correlated metal with strong quasiparticle
renormalization or even to a Mott insulating state \cite{imada98}.
Correlations may be stronger \cite{liebsch03,zhong15} or weaker
\cite{lantz15} than in the corresponding bulk state.  This kind of
physics is included in and described by the local but dynamic DMFT
correlations.  Another situation where electronic correlations are
important is a Hund's metal \cite{yin11,georges13} where several $d$
electrons form a local magnetic moment;  the Hund's exchange $J$, and
not the Coulomb repulsion $U$ plays the decisive role.  Here, the effect
of electronic correlations is less pronounced in the one-particle
spectrum, but strongly reflects in two-particle correlation functions
such as e.g.\ the magnetic susceptibility.

Electronic correlations can give rise to magnetic and/or orbital
ordering. Indeed magnetism and possible spintronic applications is one
of the prospective advantages of oxide heterostructures. Such an
ordering can already be described by the  simpler DFT+$U$
\cite{anisimov91} which is a Hartree-Fock treatment and hence does not
include genuine correlations. Indeed, a fully polarized ferromagnet is a single
Slater determinant and such a ground state is perfectly accounted for in
DFT+$U$. What DMFT additionally covers is the correlated paramagnetic
state which competes with the magnetic state. This competition is most
relevant for the question whether there is ordering or not. Hence DMFT
is superior to DFT+$U$ in its predictive power regarding ordering and
also allows us to calculate the critical temperature without further
approximations or without an adjustment of $U$.

For the PES also the excited states are important even if the ground
state is a single Slater determinant.  In this respect, DMFT e.g.\
describes the extra spin-polaron peaks \cite{sangiovanni06} in the
spectrum of an antiferromagnet which is beyond the Hartree-Fock
physics of DFT+$U$.  A symmetry broken DFT+$U$ calculation is often
employed for mimicking a localized state also in the paramagnetic
phase. An example are the localized states at oxygen vacancies
mentioned above which can be described in both,  DMFT
\cite{behrmann15} and DFT+$U$ \cite{cuong07,zhong10,jeschke15}. For
the PES, DFT+$U$  is at least a valid first approximation.  Some
aspect such as the spin-polaron peaks are missing, but a gap in the
spectrum and a localized state can be described this way.

An important aspect of magnetism is the screening of the magnetic
moment.  Even though we might have a magnetic moment on the femtosecond
time scale, this moment and its direction might fluctuate  in time
\cite{hansmann10prl}. It is screened. Obviously this is an effect of
dynamic correlations included in DMFT, but neither in DFT nor DFT+$U$.
This suppression of the long-time magnetic moment is important for
magnetism \cite{yin11,galler15}. Indeed it is one physical reason why
there is no magnetism even if it is predicted by DFT which usually
underestimates correlations and tendencies towards magnetic ordering in
transition metal oxides. Note that DFT+$U$, on the other hand, grossly
overestimates tendencies of magnetic ordering because  its only way to
avoid the interaction $U$ is ordering. The screening is a further reason
why DMFT is more reliable regarding predictions of magnetic ordering.
Let us add that experimentally the short-time local moment is
discernible in fast, e.g., x-ray absorption, experiments, whereas no
moment will be seen in experiments on a longer time scale, e.g., when
measuring the magnetic susceptibility.

DMFT includes local dynamic (quantum) correlations, but neglects spatial
correlations. Such non-local correlations can be more important  at
lower temperatures and for lower dimensional systems.  Thence they may be
also important for oxide heterostructures which are intrinsically
two-dimensional. Non-local correlations give rise to additional physical
phenomena beyond the realm of DMFT; and one should always be aware of
the limitations of the method employed.  Physical phenomena that might
be relevant for oxide heterostructures and rely on such non-local
correlations are: excitons and further vertex corrections to the
conductivity such as weak localization, spin fluctuations that might
suppress the DMFT-calculated critical temperature for magnetic ordering,
and unconventional superconductivity. Often such phenomena can be
understood in terms of simple ladder diagrams in orders of $U$, a
treatment which is however not sufficient if correlations are truly
strong. Cluster \cite{maier05} and diagrammatic extensions
\cite{kusunose06,toschi07,rubtsov08,slezak09,katanin09} of DMFT which
include all the local DMFT correlations but also non-local correlations
beyond are a promising way to include such effects. The latter diagrammatic
extensions describe a similar physics understood by the aforementioned
ladder diagrams at weak coupling, but now for a strongly correlated
system including all the DMFT correlations, e.g., the quasiparticle
renormalization.  These non-local approaches might, in the
future, help us understand non-local correlations, but for the time
being DMFT is state-of-the-art for correlations in oxide
heterostructures and it will remain the method of choice whenever local
correlations play the dominant role.

In this chapter, we review the DFT+DMFT approach and its application to
oxide heterostructures. Section~\ref{Section:DFTDMFT} is devoted to
methodological aspects.  After explaining the advantages of a DFT+DMFT
treatment (Section~\ref{Section:DMFTmotivation}), we guide the reader
through the main steps of a DFT+DMFT calculation for a typical
correlated heterostructure --- a bilayer of SrVO$_3$ on a SrTiO$_3$
substrate (Section~\ref{Section:DMFTstepbystep}).  We also briefly
discuss how to DFT+DMFT results can be compared with the experimental
spectra (Section~\ref{Section:comparison}).

Section~\ref{Section:applications} reviews the results obtained so far
by DFT+DMFT for oxide heterostructures, focusing on titanates,
nickelates, vanadates and ruthenates.  We start  in
Section~\ref{Section:titanates}  with titanates, for which electronic
reconstruction can lead to a metallic interface but also  oxygen
vacancies are relevant as a competing mechanism and can give rise to
localized states.  In Section~\ref{Section:nickelates} we turn to
nickelates which were the first oxide heterostructure studied in
DFT+DMFT. Here,  heterostructuring might give rise to a cuprate-like
Fermi surface or topological states depending on the direction of
stacking. Results for vanadates which hint at possible applications of
oxide heterostructures as solar cells or as Mott transistors follow in
Section~\ref{Section:vanadates}. Section~\ref{Section:ruthenates} is
devoted to ruthenates which are arguably most promising for
ferromagnetism and spintronic applications.  Finally
Section~\ref{Section:conclusion} summarizes the chapter and provides a
brief outlook.

\section{\label{Section:DFTDMFT} Steps of a DFT+DMFT calculation
illustrated by SVO/STO heterostructures} 
\begin{figure}[t!]
\includegraphics[width=\columnwidth]{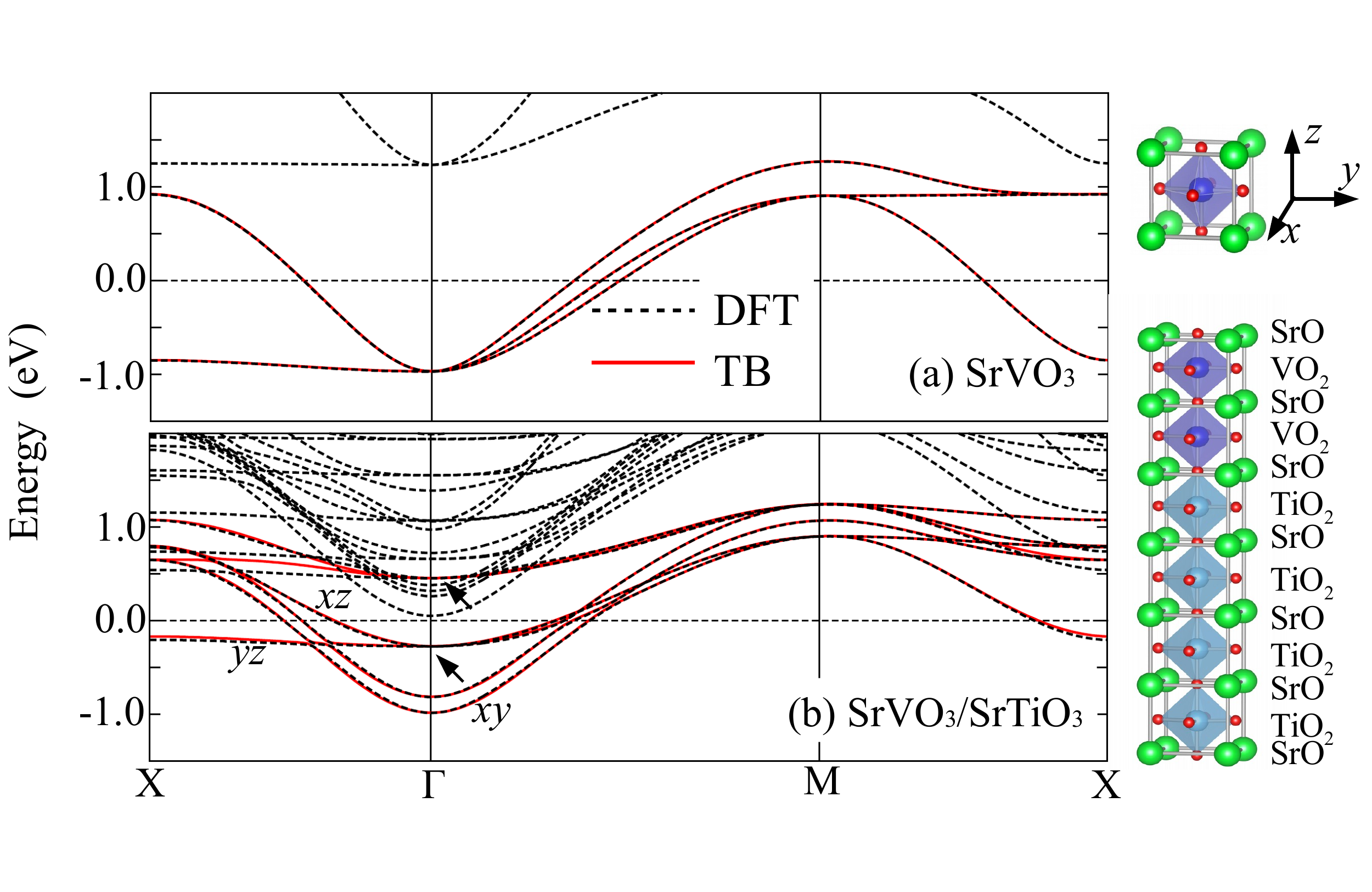}
\caption{Left: DFT band structure (black dashed lines) and the V
$t_{2g}$ bands (red solid lines) calculated for bulk SrVO$_3$. Right
top: SrVO$_3$ perovskite cell with V (blue), Sr (green) and O (red)
atoms are shown on the right.  Right bottom:   SrVO$_3$/SrTiO$_3$
heterostructure containing two SrVO$_3$ layers on a substrate of four
SrTiO$_3$ and 10{\AA} of vacuum.  The spatial confinement along $z$
leads to quantized energy levels indicated by arrows for the vanadium
$yz$/$xz$ orbitals at $\Gamma$ in the left panel.
\label{Fig1}}
\end{figure}

To illustrate the DFT+DMFT
method, we select a numerically tractable, yet instructive correlated
compound --- the cubic perovskite SrVO$_3$ (Fig.~\ref{Fig1}). Being a
rare example of a metallic V$^{4+}$ compound, it shows distinct
fingerprints of a correlated metal: the pronounced lower Hubbard peak
observed in PES~\cite{sekiyama04}, the quasiparticle peak seen in
ARPES~\cite{yoshida05}, and the upper Hubbard peak in x-ray
absorption~\cite{inoue94}.  From the computational viewpoint, the high
crystal symmetry and the $d^1$ electronic configuration render SrVO$_3$
a convenient material for testing new numerical techniques.

The strength of electronic correlations can be drastically affected by a
 dimensional reduction, as it is indeed the case for ultrathin
SrVO$_3$ layers grown on SrTiO$_3$.  Here, the reduction of the SrVO$_3$
film thickness down to two monolayers leads to a metal-insulator
transition~\cite{yoshimatsu10} and the formation of quantum well states
with an anomalous effective mass~\cite{yoshimatsu11}. In this chapter,
we will use such a 2\,SVO\,:\,4\,STO heterostructure, consisting of a
SrVO$_3$ bilayer on four SrTiO$_3$ substrate layers, as a model system
to guide the reader through the main steps of a DFT+DMFT calculation.

\subsection{\label{Section:DMFTmotivation} Motivation for DFT+DMFT: the
electronic structure of bulk SrVO$_3$} Before introducing the method, we
briefly explain the advantages of DFT+DMFT over conventional DFT
techniques and DFT+$U$.  To this end, we consider bulk SrVO$_3$.  V
atoms in its crystal structure are located in $1b$ Wyckoff positions
with the point group symmetry $m\bar{3}m$.  For $d$ electrons
($l$\,=\,2), the irreducible representations are $E_g$ (twofold
degenerate) and $T_{2g}$ (threefold degenerate).  The electrostatic
repulsion between the V $d$ and O $p$ electrons (the crystal field)
pushes the $e_{g}$ orbitals higher in energy, and the single $d$
electron of V is distributed over the three degenerate $t_{2g}$
orbitals.  

\begin{figure}[t!]
\includegraphics[width=\columnwidth]{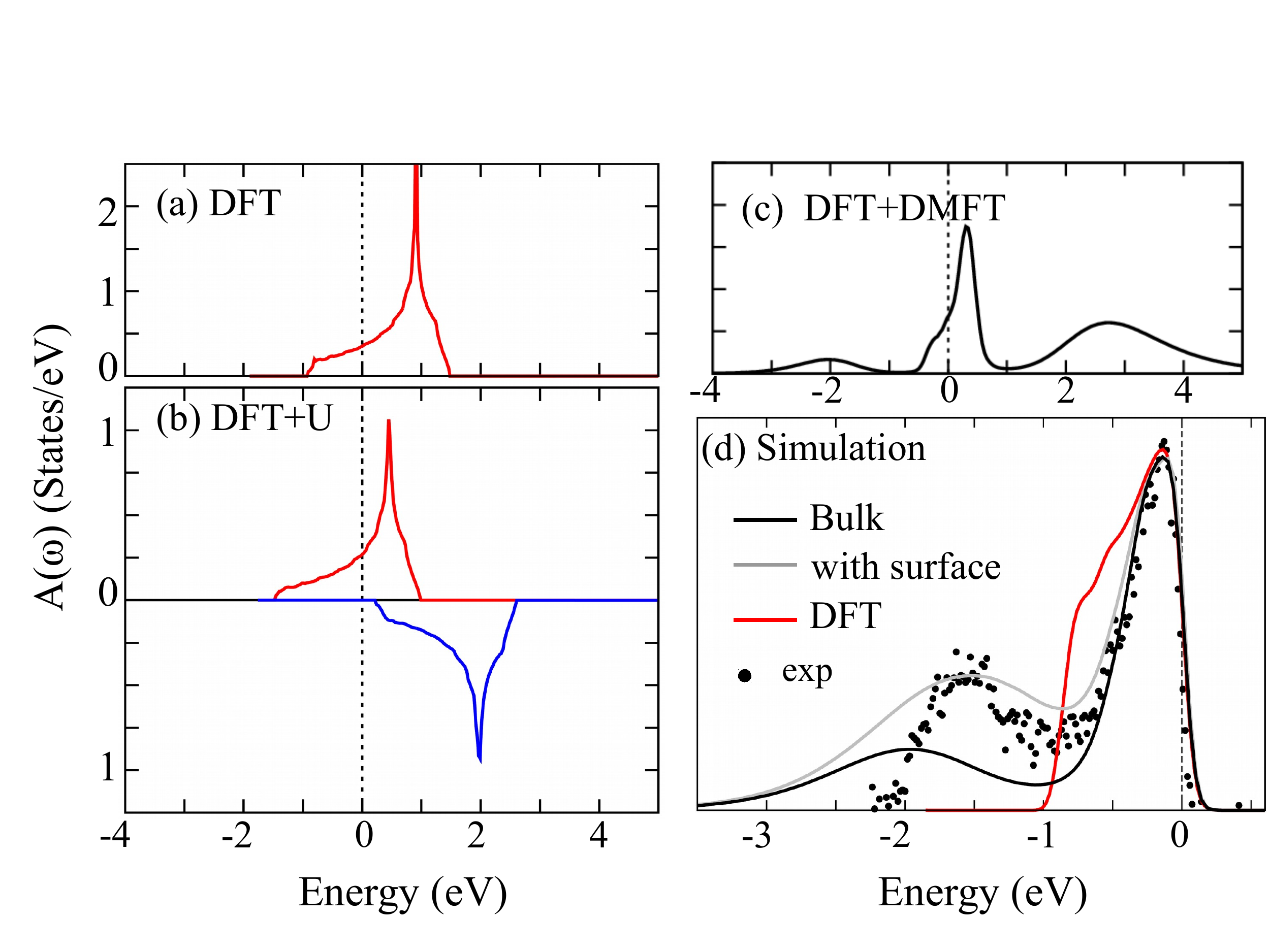}
\caption{V $t_{2g}$ DOS in (a) DFT yielding a
nonmagnetic ground state, (b) DFT+$U$ ($U_d$\,=\,2\,eV) yielding a
half-metallic magnetic state with half-filled majority states (red), and
(c) DFT+DMFT yielding a nonmagnetic solution with lower Hubbard peak,
the upper Hubbard band, and the quasiparticle peak.  The  DFT+DMFT
results show good agreement with the experiment in (d), which can be
further improved by including    SrVO$_3$ surface spectral contributions
(grey).
\label{Fig2}}
\end{figure}

The DFT band structure of bulk SrVO$_3$ is shown in Fig.~\ref{Fig1}~(a).
At the $\Gamma$ point, the three $t_{2g}$ bands are degenerate in accord
with the space group representation.  For an arbitrary $k$-point, e.g.\
on the $\Gamma$-X ($\pi/a$,0,0) path, this degeneracy is partially
lifted, and one of the bands (corresponding to the $yz$ orbital along
$\Gamma$-X) shows a minute dispersion ($\sim$0.12\,eV), whereas the two
degenerate bands ($xy$ and $xz$ orbitals) have a sizable dispersion of
$\sim$1.9\,eV since the orbital lobes are extended in the $x$ direction.
If we integrate the DFT bands over the Brillouin zone to obtain the
density of states (DOS), it becomes clear that DFT fails to describe the
experimental spectral features: the DFT DOS in Fig.\ref{Fig2} (a) lacks
the Hubbard bands, whereas the width of the quasiparticle peak is
considerably overestimated (Fig.\ref{Fig2}, d). 

The simplest DFT-based scheme which accounts for on-site interactions in
a static mean-field way is DFT+$U$.  By construction, this method favors
integer orbital occupations~\cite{ylvisaker09} and therefore is
particularly suitable for orbitally-ordered insulators, but its
applicability to correlated metals is at best limited~\cite{petukhov03}.
Indeed, for SrVO$_3$ DFT+$U$ not only fails to describe the band
renormalization, but even yields a spurious magnetic ground state
(Fig.\ref{Fig2}, b).  

The idea of DFT+$U$ is to add correlation effects to the DFT on the
Hartree-Fock level.  In this way, the DFT Hamiltonian is supplemented
with a purely real and constant self-energy $\Sigma$ for V $d$ states.
DMFT may be seen as a dynamical version of DFT+$U$, i.e.\ it accounts
also for local scattering processes, and the self-energy $\Sigma$ in
DMFT becomes complex and frequency-dependent.  This describes
(temperature-dependent) scattering processes, which lead to a finite
life time.  Thus, instead of the DOS, the spectrum of an interacting
system is described by the spectral function $A(k, \omega)$.  The
room-temperature DFT+DMFT spectral function for bulk SrVO$_3$ is shown
Fig.\ref{Fig2} (c).  By comparing it with the experimental spectra
Fig.\ref{Fig2} (d), we find a considerable improvement over DFT: all
spectral features are reproduced, and the width of the quasiparticle
peak is in good agreement with the ARPES data.

\subsection{\label{Section:DMFTstepbystep} Workflow of a DFT+DMFT
calculation}
The main steps of a DFT+DMFT calculation are\footnote{For further
details and the theoretical background we refer the reader to
\cite{held07}.}: (i) construction of the unit cell, (ii) a DFT
calculation, (iii) construction of the low-energy Hamiltonian, (iv)
mapping the lattice Hamiltonian onto a set of single-site impurity
problems (DMFT), (v) a numerical solution of the resulting single-site
models, (vi) adding the double-counting correction, and (vii) a
necessary postprocessing to compute the observables.  The way these
basic blocks are combined into a computational scheme depends on the
implementation and the convergence criterion.  In the standard DFT+DMFT
scheme, the step (ii) is repeated to converge the electronic density on
the DFT level and the steps (v) and (vi) are iterated until the
self-energy is converged~\cite{held07}.  In charge-self-consistent
calculations, the sequence of steps (ii)--(vi) is repeated until charge
redistributions become small~\cite{lechermann06,granas12,park14}.  Most
computationally demanding schemes involve structural relaxations: the
charge redistributions yielded by DMFT are used to calculate forces and
provide a new structural input, going back to the step
(i)~\cite{leonov14}.  Such a calculation terminates once the changes in
the crystal structure become small enough.

Next, we illustrate all steps in more detail using as an example two
SrVO$_3$ layers on top of a SrTiO$_3$ substrate (modeled by four
layers), following Ref.~\cite{zhong15}. The starting point of a
DFT+DMFT calculation is the construction of a unit cell.  For (001)
heterostructures, it is convenient to think of the SrVO$_3$
perovskite structure as a periodic alternation of SrO and VO$_2$
monolayers.  By stacking such monolayers on top of each other, we
construct a unit cell with two VO$_2$ layers, four TiO$_2$ layers,
and seven SrO layers.  Both terminal layers are SrO monolayers,
which ensures that transition metal atoms (V and Ti) have an
octahedral coordination.  Finally, we make the system effectively
two-dimensional by adding a sufficiently thick (10\,{\AA}) vacuum
layer along the $z$ direction, leading to an elongated unit cell
shown in the right panel of Fig.\ref{Fig1} (b).  The lateral lattice
constant $a$ of this tetragonal cell is fixed to that of the
SrTiO$_3$ substrate ($a$\,=\,3.92\,\r{A}), while all the internal
coordinates are optimized in DFT.  This latter step is crucial,
because surface and strain effects can have a big impact on the
crystal structure.  For instance, a DFT relaxation of
2\,SVO\,:\,4\,STO yields an inward drift of the surface Sr atom by
$\sim$0.25\AA. 

The next step is a DFT calculation. Here, we consider the simplest
DFT+DMFT scheme: DFT convergence is reached before going to the next
step.  At present, a plethora of very accurate DFT codes
exists~\cite{lejaeghere16}; here, we choose the all-electron full
potential augmented plane-wave method implemented in
Wien2k~\cite{schwarz02}. It is important to keep in mind that DFT+DMFT
results can in some cases sensibly depend on the chosen DFT functional.
Most DFT+DMFT studies employ either the local density approximation
(LDA)~\cite{kohn65,perdew92} or the generalized gradient approximation
(GGA)~\cite{perdew96}; here we will use the latter.  Finally, DFT
calculations are performed on a $k$-mesh, and heterostructures have the
advantage of the reduced dimensionality. DFT calculations presented here
are performed on a 10$\times$10$\times$1 $k$-mesh.

The GGA band structure of the 2\,SVO\,/\,4\,STO heterostructure is shown
in Fig.~\ref{Fig1} (b).  The thicket of bands above the Fermi energy are
mostly unoccupied Ti $t_{2g}$ states that will be neglected later in the
low-energy model~\cite{zhong12}. Moreover, the unit cell has now two
inequivalent VO$_2$ monolayers, which doubles the number of V bands.
But the main difference is that the electronic structure becomes
two-dimensional: The insulating SrTiO$_3$ substrate on one side and the
vacuum on the other side confine the V $d$ electrons to move in the
plane of the bilayer.  The partially broken translational symmetry
reduces the V point group to $4mm$, and the $t_{2g}$ bands at $\Gamma$
split into $a_{1g}$ ($xy$) and $e_{g}'$ ($yz$ and $xz$) manifolds.
Moreover, the dissimilar potential felt by V atoms in the surface and
subsurface VO$_2$ layers lifts the degeneracy between the V sites.

DFT results can be used to construct a low-energy model and
parameterize its tight-binding part.  To this end, the DFT
Hamiltonian $H^{\rm{DFT}}$ is projected onto a set of localized
orbitals. The choice of this correlated subspace depends on the
nature of the compound as well as on the problem at hand.  For
instance, to describe the spectrum of a charge-transfer insulator,
one needs to include ligand $p$ states.  But in the case of
SrVO$_3$, a natural choice for the minimal model is to select the
subspace of V $t_{2g}$ orbitals.   Since there are two inequivalent
V atoms in the 2\,SVO\,/\,4\,STO heterostructure, the number of
orbitals, and hence the dimension of the Hamiltonian matrix are
doubled.  Using this basis, we search for a tight-binding
Hamiltonian $H^{\text{TB}}$, which reproduces the DFT band structure
in the corresponding energy range.  We therefore construct
V-centered maximally localized Wannier orbitals
\cite{marzari97,marzari03, marzari12} using the Wannier90
code~\cite{mostofi08} and the wien2wannier~\cite{kunes10} interface
which has been integrated into Wien2K recently.

From the maximally localized Wannier functions, we obtain matrix
elements of the 6$\times$6 $H^{\text{TB}}(\vec{k})$ matrix that have the
form $\sum_{ij}{t_{ij}\exp{(i\vec{R}_{ij}\cdot\vec{k})}}$, where
$t_{ij}$ is the transfer integral (hopping) between the orbitals $i$,
and $\vec{R}_{ij}$ connects their centers.  The leading $t_{ij}$ terms
are listed in Table~\ref{Table:tij}.  The main effect of the confinement
is the drastic reduction of the onsite energy for the V $xy$ orbital of
the surface layer, ensuing from the surface reconstruction (the inward
drift of surface Sr atoms). 
 
\begin{table}[tb]
\caption{\label{Table:tij}
Leading transfer integrals $t_{ij}(\vec{R}_{ij})$  (in meV) for bulk SVO
and  a 2\,SVO\,/\,4\,STO heterostructure: onsite terms (first row) and
nearest neighbors hopping (along $x$; second row).  Due to the mutual
orthogonality of $t_{2g}$ orbitals, only neighboring terms with the same
orbital character are nonzero.}

\begin{tabular}{rrrrp{.02\textwidth}rrrp{.02\textwidth}rrr}
 & \multicolumn{3}{l}{bulk SrVO$_3$ } & &
\multicolumn{7}{c}{2\,SVO\,/\,4\,STO heterostructure} \\
& & & & & \multicolumn{3}{l}{surface V}
& & \multicolumn{3}{l}{subsurface V} \\
 transfer integral  &  
\multicolumn{1}{c}{$xy$} & \multicolumn{1}{c}{$yz$} & \multicolumn{1}{c}{$xz$} & &
\multicolumn{1}{c}{$xy$} & \multicolumn{1}{c}{$yz$} & \multicolumn{1}{c}{$xz$} & &
\multicolumn{1}{c}{$xy$} & \multicolumn{1}{c}{$yz$} & \multicolumn{1}{c}{$xz$} \\
\hline
onsite energies ($|\vec{R_{ij}}|$\,=\,0) &
579 & 579 & 579 & &
399 & 576 & 576 & &
540 & 574 & 574\\
nearest-neighbors along $x$ ($\vec{R_{ij}}$\,=\,$\vec{a}$)  &
$-259$ & $-26$ & $-259$ & &
$-243$ & $-37$ & $-180$ & &
$-241$ & $-33$ & $-262$\\
\end{tabular}
\end{table}

$H^{\text{TB}}(\vec{k})$ describes transfer processes of the low-energy
model.  In the spirit of the Hubbard model, we supplement this
Hamiltonian with an onsite interaction term.  In the general case, the
onsite interaction is described by the interaction vertex $U_{lmno}$,
where $l$, $m$, $n$, and $o$ and orbital indices.  For practical
purposes, simplified versions of this form are typically used~\cite{georges13},
e.g.\ the reduction to the density-density interaction $[$$l=m$ (same spin),
and $n=o$ (same spin)$]$.  Here, we use the rotationally invariant form given
by Kanamori ($l=m$ and $n=o$), which in addition to density-density
interactions, accounts accounts for spin exchange and pair hopping processes.
The Kanamori Hamiltonian has two parameters: the intra-orbital Coulomb
repulsion $U$ and the Hund's exchange $J$, while the inter-orbital Coulomb
repulsion $U'$\,=\,$U-2J$ follows from the symmetry.  For V $d$ orbitals, we
adopt $U'$\,=\,3.55\,eV from constrained LDA calculations~\cite{sekiyama04,
nekrasov06} and $J$\,=\,0.75\,eV, which is a standard value for early $3d$
metal oxides.

The resulting Hamiltonian represents a multi-orbital Hubbard model on a
two-dimensional lattice.  DMFT provides an approximate solution of it,
by performing a mapping onto a set of single-site Anderson impurity
problems.  The unit cells of heterostructures often consist of many (two
in the case of 2\,SVO\,/\,4\,STO) correlated atoms, and therefore this
mapping involves several steps.  First, we denote with $n$ and $m$ the
number of atoms in the cell and the number of orbitals per atom,
respectively (for the SVO bilayer and V $t_{2g}$ orbitals we have
$n$\,=\,2 and $m$\,=\,3).  Depending on the number and on the type of
Wannier functions chosen in the downfolding, the $m$ orbitals associated
to each atom can be either considered all as correlated (e.g. the three
$t_{2g}$-orbitals here) or can be further split in a subset of
correlated ones and a subset of ``ligands'' (e.g. $p$-orbitals).  In the
latter case the ligands are formally associated to one of the $n$ atoms
in the cell, but this is only an arbitrary assignment.  In the standard
treatment, the interaction term affects only the $d$ correlated orbitals
and it is applied locally to each of the $n$ atoms.

We first define the local Green's function
$G_{\text{loc}}$:

\begin{equation}
\label{Eq:Gloc}
G_{\text{loc}}(\omega) = \sum_{\vec{k}}\left[\left(\omega +
\mu\right)\mathbf{I} - H(\vec{k}) -
\Sigma(\omega) - H_{\text{DC}}\right]^{-1},
\end{equation}

where $\omega$ is the complex frequency, $\mu$ is the chemical potential,
$\mathbf{I}$ is the unitary matrix and $\Sigma(\omega)$ the self energy 
matrix  of dimension  $mn\times{}mn$. The latter is usually set to
zero in the first DMFT cycle, and $H_{\text{DC}}$ is the double-counting
correction, which will be discussed later.  

We now consider the $i$-th atom and start to construct the corresponding
impurity problem.  The $i$-th $m\times{}m$ block along the diagonal of
$G_{\text{loc}}$ is denoted as $G_{\text{loc}, i}$
(0\,$<$\,$i$\,$\leq$\,$n$).  The $i$-th impurity Weiss' field
$\mathcal{G}_{i}$ is constructed by inverting the corresponding
$G_{\text{loc}, i}$:
\begin{equation}
\label{Eq:Gi}
\mathcal{G}_{i}(\omega) = \left[(G_{\text{loc}, i})^{-1} + \Sigma_i(\omega)
\right]^{-1},
\end{equation}
where $\Sigma_i(\omega)$ is now a $m\times{}m$ diagonal matrix
containing the self-energy of that block.  It is important to note
that $\mathcal{G}_{\text{loc}, i}(\omega)$ contains nonlocal
contributions stemming from the off-block-diagonal elements of
$H^{\text{TB}}$, that enter $G_{\text{loc}}$ (and hence
$G_{\text{loc}, i}$) by matrix inversion in Eq.~\ref{Eq:Gloc}.

The impurity Green's function $\mathcal{G}_{i}(\omega)$ allows us to
formulate the corresponding Anderson impurity problem, amenable to a
numerical solution.  Various techniques can be
used~\cite{georges96,held07}, yet presently the method of choice is the
continuous time quantum Monte Carlo in the hybridization expansion
(CT-HYB, see~\cite{gull11} for a review; an implementation for the
Kanamori Hamiltonian is discussed in~\cite{parragh12}).  The main
parameters of a CT-HYB calculation are the inverse temperature $\beta$
($\beta$\,=\,39.6\,eV$^{-1}$ corresponds to room temperature), the
interaction parameters and the number of Monte Carlo sweeps.  After the
CT-HYB calculations ($n$ of them or less, in case some of the atoms are
locally equivalent) we access the set of self-energies
$\Sigma_{i}(\omega)$ (0\,$<$\,$i$\,$\leq$\,$n$).  This in turn allows
for the construction of a new local Green's function $G_{\text{loc}}$
for the whole heterostructure, as in Eq.~\ref{Eq:Gloc}.  Let us stress
again that in DMFT the self-energy is frequency-dependent, but does not
depend on $\vec{k}$, because the impurity model is a local problem.
With the new Green's function, we can start another DMFT cycle.  This
procedure is repeated unit convergence.

The only term that we have not discussed so far is the double-counting
correction $H_{\text{DC}}$ (Eq.~\ref{Eq:Gloc}).  The underlying idea is
to subtract those contributions to $\Sigma_{i}$ that are already
accounted for by the GGA.  Unfortunately, GGA lacks a diagrammatic
description, and such contributions can not be determined rigorously.
Several forms of $H_{\text{DC}}$ have been proposed, but no general
solution to this conundrum exists, a problem also  relevant
for DFT+$U$.  However, in our calculation with $t_{2g}$-orbitals only
the symmetry is close to cubic  where  $H_{\text{DC}}$ 
is equivalent to a trivial shift of the chemical potential $\mu$~\cite{held07}.

Several remarks concerning the performance should be made.  For typical
(not too large) supercells, the computationally most time-consuming step
is solving the DMFT impurity problems.  Here, the crucial parameter is
the number of orbitals ($m$), because for CT-HYB the computational time
scales exponentially with $m$.  In contrast, the number of sites ($n$)
defines the number of impurity problems to solve, and the computational
time scales only linearly with $n$. For larger supercells the matrix
inversion in Eq.\ (\ref{Eq:Gloc}) which  scales like $(mn)^3$ (or best
$\sim (mn)^{2.373}$) becomes the bottleneck.

\subsection{\label{Section:comparison} Comparison with the experiment}
The final step of a DFT+DMFT calculation is postprocessing.  CT-QMC
calculates Green's functions $G(i\omega)$ on the imaginary frequency
(Matsubara) axis and an analytical continuation is required in order to
compare with the experimental $A(\omega)$ spectra.  Although there
exists a one-to-one correspondence between $G(\omega)$ and $G(i\omega)$,
a straightforward solution is practically impossible.  The root cause of
the problem are statistical errors in $G(i\omega)$ that give rise to
huge differences in the analytically-continued $G(\omega)$.  A standard
solution to this problem is the maximum entropy method~\cite{jarrell96}.

After an analytical continuation of the self-energy, the spectral
function $A(\vec{k}, \omega)$ can be evaluated as
\begin{equation}
\label{Eq:Awk}
A(k, \omega)=-\frac{1}{\pi}\text{Im}\,G(k,
\omega)=-\frac{1}{\pi}\text{Im}{\left[\left(\omega +
\mu\right)\mathbf{I}_m\otimes\mathbf{I}_n - H(k) -
\Sigma(\omega)\right]^{-1}}.
\end{equation}
Orbital occupations can be obtained by integrating the respective
diagonal elements of $A(\vec{k}, \omega)$ over the Brillouin zone in the
negative frequency range up to the Fermi energy.

\begin{figure}[t!]
\includegraphics[width=\columnwidth]{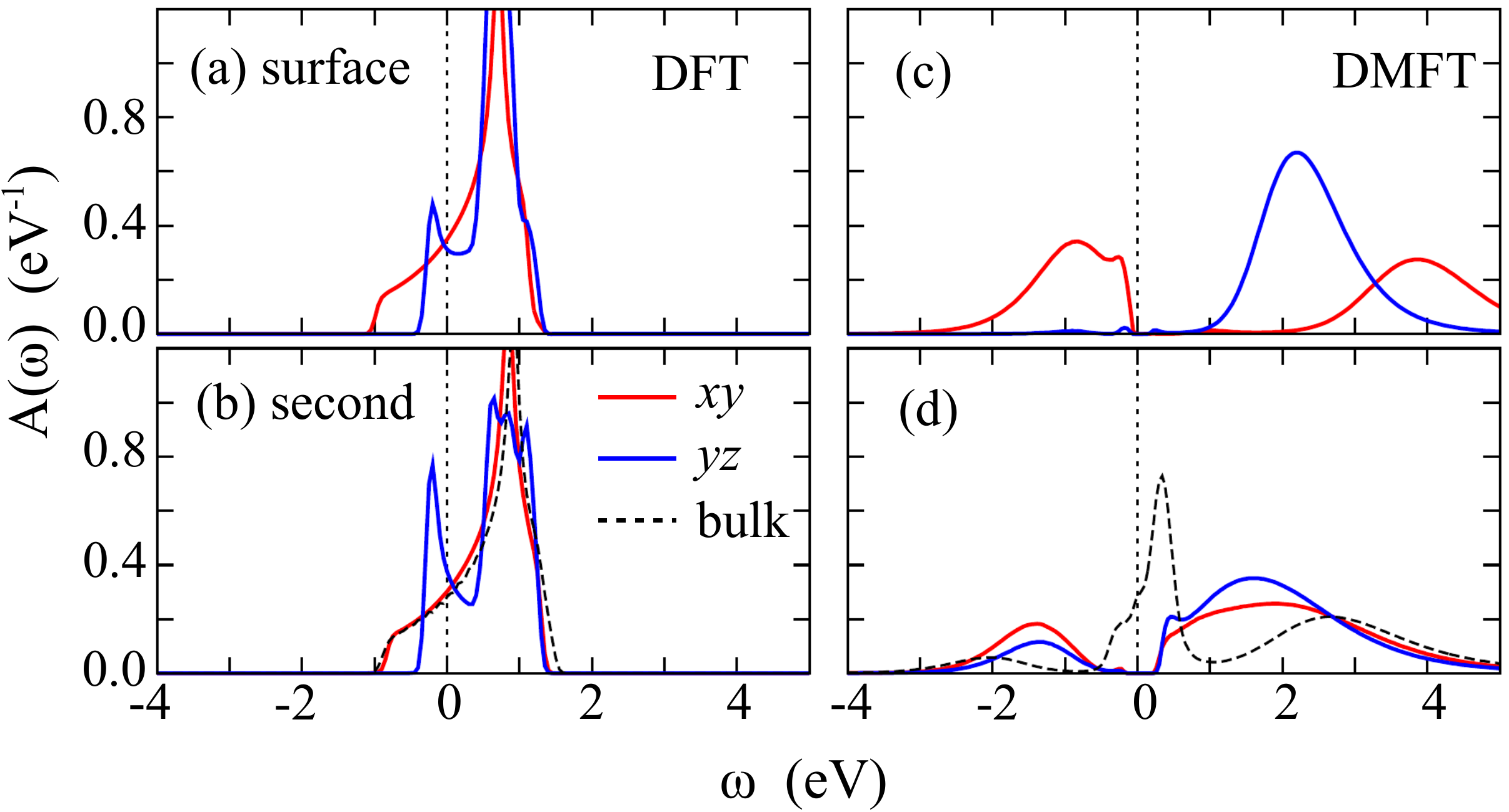}
\caption{Layer- and orbital-resolved DOS (left) and the respective
DFT+DMFT spectral functions (right) of the 2\,SVO\,/\,4\,STO
heterostructure, compared to bulk SrVO$_3$ (dashed
line) (adapted from \cite{zhong15}).
\label{Fig:Awk}}
\end{figure}

We are now in the position to compare the DFT+DMFT spectral functions
with the DOS and in this way evaluate the effect of electronic
correlations.  The left panels of Fig.\ref{Fig:Awk} show the layer- and
orbital-resolved DOS for the 2\,SVO\,/\,4\,STO heterostructure, with the
nonzero DOS at the Fermi level indicating that both layers are metallic.
The  $xy$ DOS is similar to bulk SrVO$_3$, but for  the $yz/xz$ orbitals
the confinement in the $z$ direction leads to two peaks in the
Fig.~\ref{Fig:Awk} (a, b) and a narrowing of the DOS.  The  $yz/xz$ DOS
center of mass is shifted upwards to higher energies compare the  onsite
terms in Table \ref{Table:tij}.   Fig.~\ref{Fig:Awk} (right) shows the
room-temperature DFT+DMFT spectral functions for 2\,SVO\,/\,4\,STO. The
effect of correlations is dramatic \cite{zhong15}: In the surface layer,
they trigger the orbital polarization which renders the $xy$ orbital
half-filled.  At half-filling, the intraorbital Coulomb repulsion is
particularly efficient, and it splits the spectrum into lower and upper
Hubbard bands, stabilizing the Mott insulating state.  The subsurface
layer also becomes insulating, albeit with a weaker orbital polarization
and a much narrower gap.  

\begin{figure}[t]
\includegraphics[width=\columnwidth]{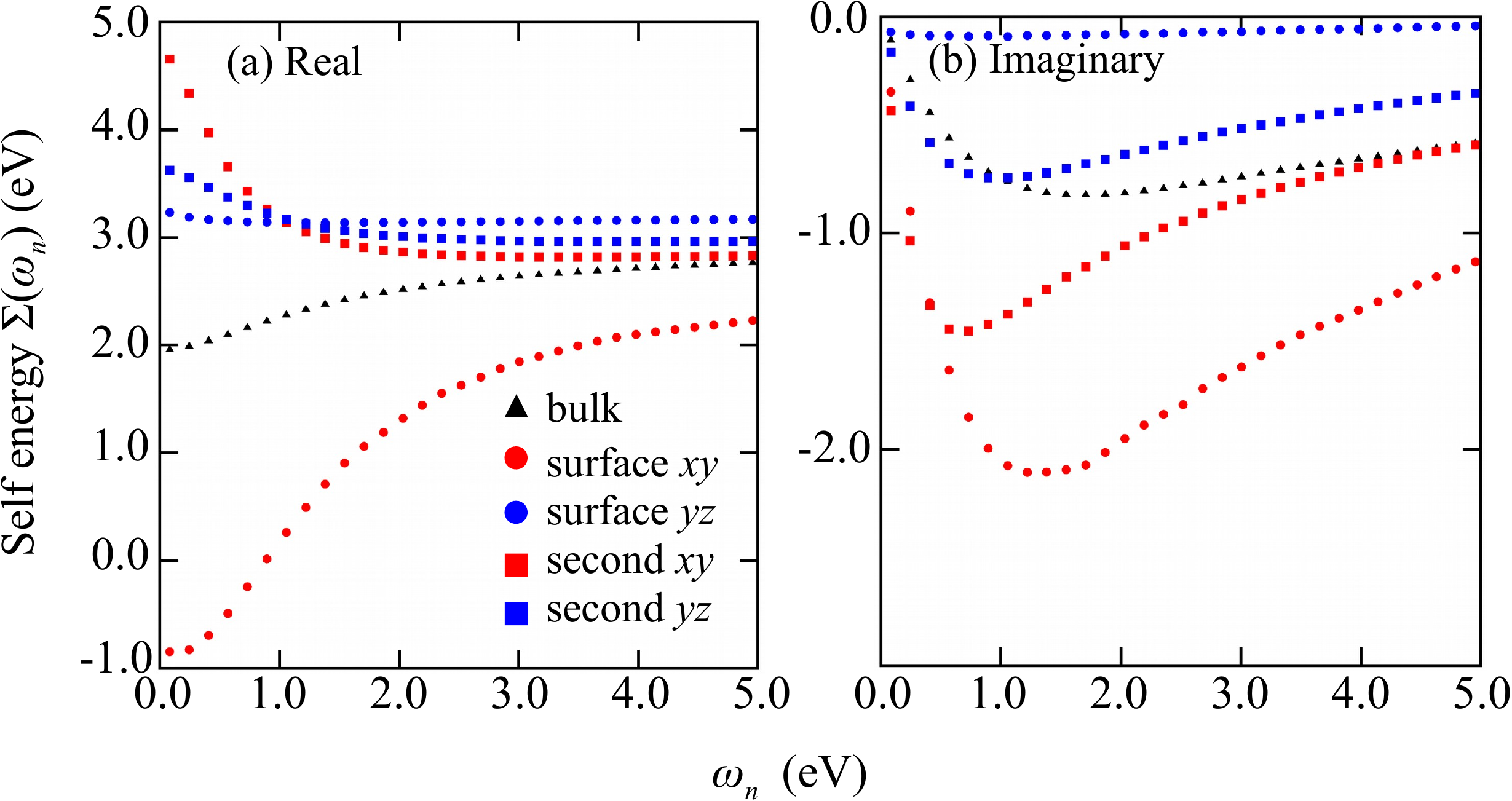}
\caption{(a) Real and (b) imaginary part of the layer- and
orbital-resolved DMFT self-energy for 2\,SVO\,/\,4\,STO
 as a function of real frequency $\omega$. Results for
bulk SrVO$_3$ are shown to illustrate the effect of dimensional
reduction.
\label{Fig:sigma}}
\end{figure}

Fig.~\ref{Fig:sigma} shows the  self-energy $\Sigma(\omega)$ of the
2\,SVO\,/\,4\,STO heterostructure.  Correlations in the $xy$ orbital are
particularly strong, as reflected in the sizable frequency dependence of
$\text{Re}\,\Sigma$. At higher frequencies, the DMFT self-energy
recovers the static Hartree shift, but it is the low energy part of  of
$\text{Re}\,\Sigma$ which deviates strongly for the different orbitals,
enhances the DFT crystal field splitting and leads to insulating SVO
layers, similar as in \cite{Keller2004}.  

From the metallic-like behavior of $\text{Im}\,\Sigma$, we can conclude
that the  2\,SVO\,/\,4\,STO heterostructure is on the verge of a Mott
transition. Switching between metallic and insulating regimes can be
achieved by strain, doping, applying external electric field, or capping
the surface layer. DFT+DMFT is an excellent computational tool to
explore these possibilities and evaluate the optimal way to tune the
physics of this interesting system~\cite{zhong15}.

\section{\label{Section:applications} Applications}
After the discovery of the emergent metallicity at the
LaTiO$_3$/SrTiO$_3$ interfaces, one of the major challenges for theory
was to explain the mechanism underlying the formation of the metallic
state at the interface between  a band insulator (SrTiO$_3$) and a Mott
insulator  (LaTiO$_3$).  This dichotomy motivated Okamoto and Millis to
build up the concept of ``electronic reconstruction'', extending
earlier ideas for fullerenes~\cite{hesper00} to oxides.  The first model
calculations~\cite{okamoto04nat,okamoto06} along these lines included
correlation effects on the Hartree-Fock level.  A DMFT model calculation
for the LaTiO$_3$/SrTiO$_3$ heterostructure
followed~\cite{okamoto04prb1}, showing a ``leakage'' of electrons from
the Mott- to the band-insulating side.  This gives rise to a partially
filled $d$-shell at the interface layer which facilitates conductivity.
Even though the low-energy model comprised a single orbital only, these
results strongly contributed to the understanding of the electronic
reconstruction mechanism.  Later realistic DFT+DMFT calculations for
oxide heterostructures followed, starting with the work of Hansmann {\em
et al.}  for nickelates \cite{hansmann09}.  In the following four
Sections we review DFT+DMFT calculations for oxide heterostructures,
which focused hitherto on titanates, nickelates, vanadates, and
ruthenates.
 
\subsection{Titanates} 
\label{Section:titanates}
Titanates are by far the most studied class of oxide heterostructures.
At the same time, they are arguably also the best understood class, and
the main ingredients of their emergent behaviors are known from detailed
DFT+DMFT studies, pioneered by Lechermann {\em et al.}
\cite{lechermann13, lechermann14, behrmann15, lechermann16}. In
particular, this is the case for LaTiO$_3$/SrTiO$_3$ (LTO/STO) where
DFT+DMFT is needed to properly describe the Mott insulator LTO.  In the
LaAlO$_3$/SrTiO$_3$ heterostructure on the other hand, which is
experimentally most frequently  studied,  LaAlO$_3$ (LAO) is instead a
simple band insulator. The metallic state induced by a slight doping due
to oxygen vacancies or  electronic reconstruction, is hence only weakly
correlated. This situation in LAO/STO can be well described by DFT,
except for states localized at the oxygen vacancies.

State-of-the-art DFT+DMFT(CT-HYB) calculations for the LTO/STO
heterostructure started with Lechermann {\it et al.}~\cite{lechermann13}
who reported the formation of a quasiparticle peak primarily of
$d_{xy}$-orbital character, due to the lifted $t_{2g}$-degeneracy and in
line with the experimental observations.  A subsequent study focused on
the magnetic properties and revealed that ferromagnetism is stabilized
by the joint effect of electronic correlations and oxygen
vacancies~\cite{lechermann14}.  The role of the latter is a question of
fundamental importance, and DFT+$U$ (see e.g. \cite{zhong10}) and
DFT+DMFT~\cite{behrmann15, lechermann16, altmeyer16} studies are at the
forefront of the research. DFT+DMFT indicates the formation of in-gap
states in LAO/STO~\cite{lechermann14,lechermann16,altmeyer16}, in good
agreement with PES measurements.  This is definitely one of the most
promising directions of investigation, even though the presence of
oxygen vacancies increases the computational effort tremendously.  The
strongly correlated nature of such defect states gives rise to a strong
sensitivity of the electronic properties to small changes in the
interaction parameters or local distortions ensuing from the structural
relaxation.

Another important aspect are local distortions and rotations of the
oxygen octahedra. Their role in stabilizing an insulating state in bulk
$d^1$ titanates has been advocated  by Pavarini {\it et
al.}~\cite{pavarini04, pavarini05}.   Heterostructuring strongly affects
these distortions and rotations because  of  strain and breaking
translational invariance.  Dymkowski and Ederer found in  DFT+DFMT that
compressive strain reduces the $t_{2g}$-splitting which is intimately
related to the collective tilts and rotations,  and eventually turns
LaTiO$_3$ into a metal~\cite{dymkowski14}. 

More recently a promising route towards new electronic behaviors,
the so-called $\delta$-doping, has been proposed, i.e., adding a
single impurity layer into a layered superlattice.  To this end,
one of the LaO layers in a large LaTiO$_3$ supercell has been
replaced by an SrO layer and studied using
DFT+DMFT~\cite{lechermann15}. It was found that depending on the
distance from the interface, titanate layers exhibit three distinct
electronic states~\cite{lechermann15}.

\subsection{Nickelates} 
\label{Section:nickelates}

Bulk rare-earth nickelates RNiO$_3$ with trivalent Ni$^{3+}$ exhibit
interesting electronic, magnetic and transport
properties~\cite{garcia-munoz92, torrance92, catalan08}.  One of the
most remarkable phenomena is the metal-insulator transition  observed in
all RNiO$_3$ materials except for LaNiO$_3$ (LNO), which remains
metallic down to the lowest temperatures~\cite{garcia-munoz92}.  The
nature of this metal-insulator transition has been addressed by
experiment~\cite{rodriguez-carvajal98, alonso99, staub02, nowadnick15}
and theory~\cite{lee11, lee11prb, lau13, nowadnick15}, but the
discussion is not yet settled.

In the simplest ionic model, Ni$^{3+}$ has a $d^7$ configuration in LNO.
The octahedral crystal field splits the $d$ levels into $t_{2g}$ and
$e_{g}$ manifolds, with   the low-lying $t_{2g}$ states completely
filled.  Hence all electronic, orbital and spin degrees of freedom
pertain to the $e_{g}$ states.  Close similarities between nickelates
and cuprates were noticed in the early days of high-temperature
superconductors.  Indeed, the $e_{g}$ shell of $d^7$ Ni with its single
electron is seemingly a sibling of $d^9$ Cu having a single hole.  Thus,
a hypothetical nickelate with half-filled $x^2-y^2$ and empty $3z^2-r^2$
orbitals would have a cuprate-like Fermi surface and eventually become
superconducting upon doping. But bulk Ni$^{3+}$ systems do not exhibit
such an orbital polarization~\cite{anisimov99}.

Chaloupka and Khaliulin~\cite{chaloupka08} suggested to resort instead
to nickelate heterostructures, with the idea in mind that the $z$ axis
confinement leads to a reduced bandwidth of the $3z^2$-$r^2$ orbital
which  hence gets Mott insulating, leaving behind a $x^2$-$y^2$ Fermi
surface. In realistic DFT+DMFT calculations,  Hansmann {\em et
al.}~\cite{hansmann09} found that it is instead the correlation enhanced
crystal field splitting between  $x^2$-$y^2$ and $3z^2$-$r^2$ orbitals
which pushes the $3z^2$-$r^2$ orbital higher in energy and gives rise to
single-sheet Fermi surface of  $x^2$-$y^2$ character. However, strain or
a  PrScO$_3$ substrates is needed to achieve this, or unrealistically
large $U$ values~\cite{hansmann09}.  This can also trigger a
metal-insulator transition which is very different for the  $x^2$-$y^2$
and $3z^2$-$r^2$ orbital~\cite{hansmann10}: While
$\operatorname{Im}{\Sigma}_{x^2-y^2}(\omega)$ diverges for
$\omega\rightarrow{}0$ as in the single-band Hubbard model,
$\operatorname{Im}{\Sigma}_{3z^2-r^2}(\omega)$ retains a seemingly
metallic-like behavior. The insulating state is instead induced by the
change in $\operatorname{Re}{\Sigma}_{3z^2-r^2}(\omega)$ relative to
$\operatorname{Re}{\Sigma}_{x^2-y^2}(\omega)$ at $\omega=0$, i.e., by
the aforementioned enhancement of the crystal field splitting.  

Later,  Han {\it el al.}~\cite{han11} challenged this picture  and
vouched for an explicit inclusion of the oxygen ligand orbitals into the
calculation.  Their study found electronic correlations to decrease the
orbital polarization disregarding the initial crystal-field splitting.
The main difference between the two DMFT studies is the choice of the
basis: Hansmann~\emph{et~al.} employed a  $d$-only calculation including
the Ni $e_{g}$ states and which only requires a minimal number of
interaction parameters.  In contrast, Han~\emph{et~al.} \cite{han11}
employed a $d$-$p$ basis, where the O $2p$ orbitals are included
explicitly.  The choice of the $d$-$p$ model is motivated by the high
oxidation state of Ni$^{3+}$, which can be unstable towards the
formation of a ligand hole (L) and the transfer of an electron to the Ni
$d$ shell.  In the $d$-$p$ calculation, which allows for such
charge-transfer processes, the strong hybridization between Ni $d$ and O
$p$ states and the Hund's exchange in the $d$-shell reduce the orbital
polarization~\cite{han11}. In principle, including  additionally the O
$p$ states should be better as it allows for charge transfer processes.
However, such calculations are very sensitive to the double counting or
$d$-$p$ interaction and a commonly accepted scheme still needs to be
established.

\begin{figure}[tb]
\includegraphics[width=\columnwidth]{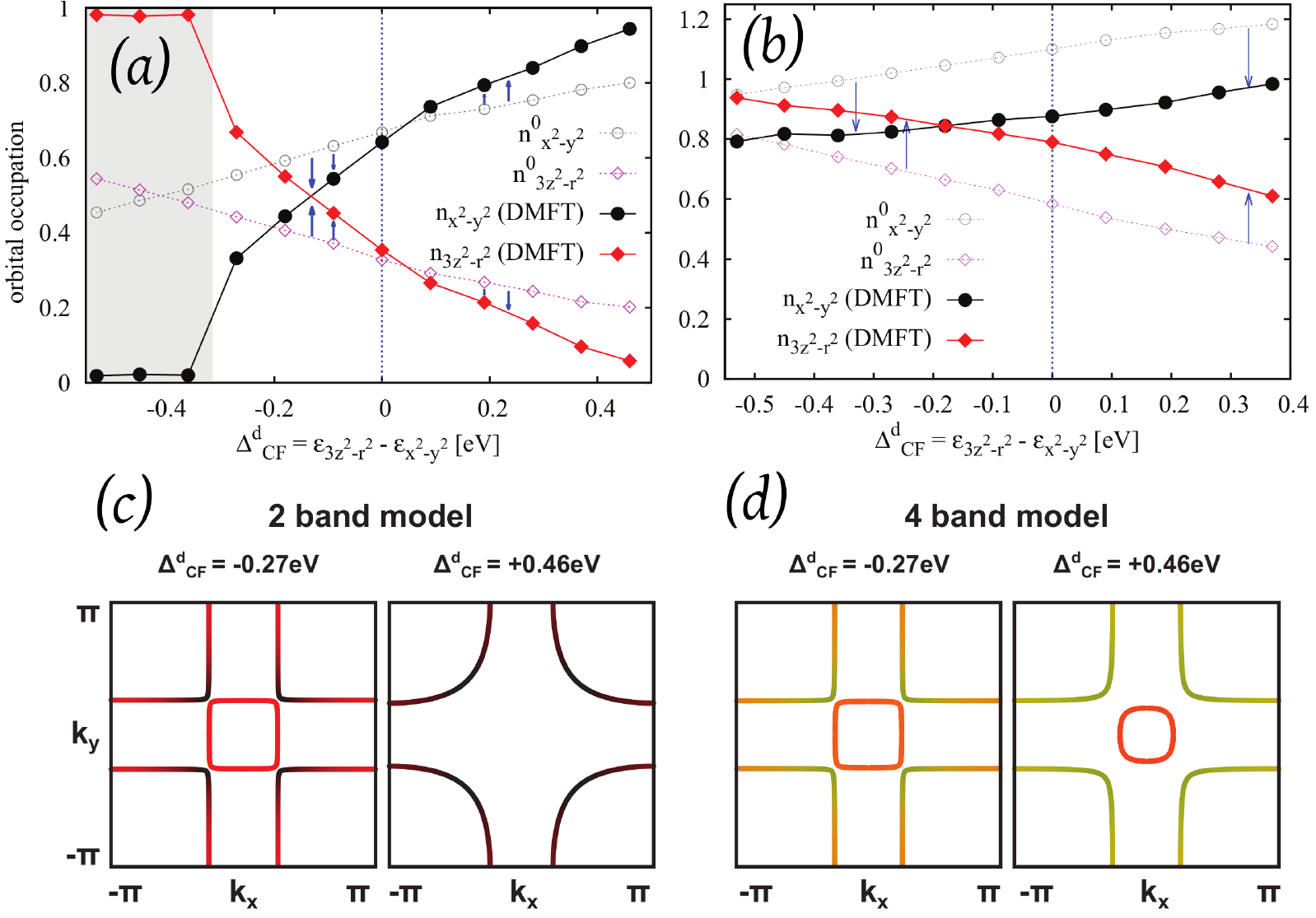}
\caption{ (a) Orbital occupations vs.\ crystal field splitting for
a $d$-only (2 band) model of a LNO monolayer without interaction (open
symbols) and within DMFT (filled symbols;  $U$\,=\,5.5\,eV,
$J$\,=\,0.75\,eV, $T$\,=\,1165\,K,  $n_e$\,=1 electron/site).  The
shaded region denotes the Mott-Hubbard insulating phase.  (b) Same for a
$d$-$p$ (4 band) model ($U$\,=\,10\,eV, $n_e$\,=\,5).  (c-d) Shape and
orbital character of the Fermi surface as calculated by DMFT for (c) the
$d$-only and (d) the $d$-$p$ model.  The color coding is as follows:
black for $x^2$-$y^2$, red for $3z^2$-$r^2$, yellow for O\,$p$.  For the
$d$-only model electronic correaltions enhance the orbital
disproportionation (a)  while for the $d$-$p$ model it is reduced,
preventing a single-sheet Fermi surface (d) as in (c).  (adapted from
\cite{parragh13}).
\label{fig:ni} }
\end{figure}

An extensive comparative DMFT study of $d$ and $d$-$p$ models  allowed
Parragh \emph{et~al.}~\cite{parragh13} to pinpoint the difference
between the two models~\cite{parragh13}, see Fig.~\ref{fig:ni}. Parragh
\emph{et~al.} found that the deviations are ruled by the filling of the
correlated $d$-shell. In the $d$-only model the filling is $d^7$ or one
electron per site in the two $e_g$ orbitals so that shifting the second
$3z^2$-$r^2$ orbital upwards and reducing its occupation is
energetically favorable (Fig.~\ref{fig:ni} a). For the $d$-$p$ model
instead an essentially $d^8$ (ligand hole) configuration brings the
Hund's exchange into the game. This favors the two $e_g$ electrons to
form a spin $S=1$, equally occupying the two different  $3z^2$-$r^2$ and
$x^2$-$y^2$ orbitals (Fig.~\ref{fig:ni} b). This actually leads to a
downward instead of an upward shift of the  $3z^2$-$r^2$
orbital~\cite{parragh13}.  This has dramatic consequences for the
topology of the Fermi surfaces: while the $d$-only model features a
single-sheet cuprate-like Fermi surface (Fig.~\ref{fig:ni} c), a second
sheet always appears in $d$-$p$ model calculations (Fig.~\ref{fig:ni}
d).  Later Peil~\emph{et~al.} applied DFT+DMFT to study strain effects
in nickelate superlattices~\cite{peil14}.  As  in Ref.~\cite{hansmann09}
they found that strain can induce a sizable orbital polarization.
However they point out that the  Hund's exchange can reduce the orbital
polarization for both the   $d^8L$  and $d^7$  configuration. The latter
$d^7$  result revises common knowledge and is an indirect consequence of
the intersite hopping. Noteworthy, their DMFT orbital polarization is
smaller than in DFT.

Which of the two calculations, $d$-only or  $d$-$p$, provides a more
appropriate framework for the fermiology and spin-state  of nickelate
heterostructures?  From the physical perspective, the $d$-$p$ model is
clearly superior to the $d$-only model: Owing to the larger energy
window used for Wannier projections, the resulting Wannier functions are
more localized and atomic-like.  In addition, the occupation of the
$d$-shell in $d$-$p$ models is flexible, and hence such models can
capture charge transfer.  However, the price to pay is the drastic
increase in the number of free parameters, including the on-site
interactions on the O $2p$ orbitals as well as $d$-$p$ interaction whose
experimental estimation is very challenging. Or, alternatively, a still
tentative double counting or an {\em ad hoc} adjustment of the  O $p$
level to  experiment is necessary.

On the experimental forefront, the situation remains controversial.  In
LAO/LNO heterostructures,  orbital polarizations (but only moderate
ones) have been found.   For 4\,LNO/\,4\,LAO heterostructures, soft
x-ray reflectivity measurements reveal a small orbital polarization of
7(4)\,$\%$ for the outer (inner) LNO monolayers~\cite{benckiser11}.  On
the other hand, by choosing a different non-correlated spacer, such as
e.g. GdScO$_3$, tan orbital polarization as high as 25\,$\%$ can be
achieved~\cite{wu13}.  Recently grown three-component  LTO/LNO/LAO
heterostructures show even larger polarizations~\cite{disa15}.  The
origin of sizable orbital polarization in these heterostructures has
been addressed in a very recent DFT+DMFT study~\cite{park16}.  In the
layered bulk nickelate Eu$_{2−x}$Sr$_x$NiO$_4$, Uchida {\em et al.}
\cite{uchida11} found a single cuprate-like Fermi surface in ARPES which
even evidences a pseudogap.  Experimentally the smoking gun would be to
determine the local magnetic moment which should be $\sim$2 and
$\sim{}1$ $\mu_{\text{B}}$  for the   $d^8L$  and $d^7$ configuration,
respectively.

\subsection{\label{Section:vanadates}  Vanadates}
In this section we review DFT+$U$ and DFT+DMFT calculations of LaVO$_3$
(LVO) heterostructures. Please note that results for SrVO$_3$/SrTiO$_3$
which might serve as a Mott transistor~\cite{zhong15}  have already been
discussed in Section \ref{Section:DMFTstepbystep}; also the charge
transfer in SrVO$_3$/SrMnO$_3$ has been analyzed~\cite{chen14}.  Bulk
LVO has been analyzed within DFT+$U$ by Fang and Nagaosa \cite{fang04}
and within DFT+DMFT by De Raychaudhury, {\it et al.}
\cite{de_raychaudhury07}.  The latter calculation revealed the existence
of strong orbital fluctuations above the N\'{e}el temperature $T_N$,
which coincides with the structural transition temperature.  These
quantum effects reduce quite rapidly the orbital polarization, in
contrast to YVO$_3$, for instance, where a pronounced orbital
polarization persists up to temperatures of the order of 1000\,K, i.e.
well above $T_N$.  The magnetism in LVO is of $C$-type, i.e.
ferromagnetically stacked planes with antiferromagnetic order within
each plane ($q$\,=\,$\pi\pi{}0$). This ordering is accompanied by an
orbital pattern of an intermediate kind between $G$- and $C$-type, due
to the competition between the Jahn-Teller of the oxygen octahedra and
the GdFeO$_3$-type distortion \cite{de_raychaudhury07}.  Note that in
SVO/LVO heterostructures also a ferromagnetic state is possible.  The
structural transition at 140\,K is from a high-temperature orthorhombic
to a low-temperature monoclinic phase.

Hotta {\em et al.} \cite{hotta07} were able to grow  LVO  epitaxially on
STO and found  a metallic $n$-type interface for thick enough LVO films,
which they interpreted to originate from the polar discontinuity.
Motivated by these experiment and also intrigued by the advantageous
size of the gap of LVO  ($\sim$1\,eV) for photovoltaic applications,
Assmann {\em et al.}   studied LVO thin films with DFT+$U$
\cite{assmann13} and  DFT+DMFT \cite{assmannphd}. 

\begin{figure}[t]
\includegraphics[width=\textwidth]{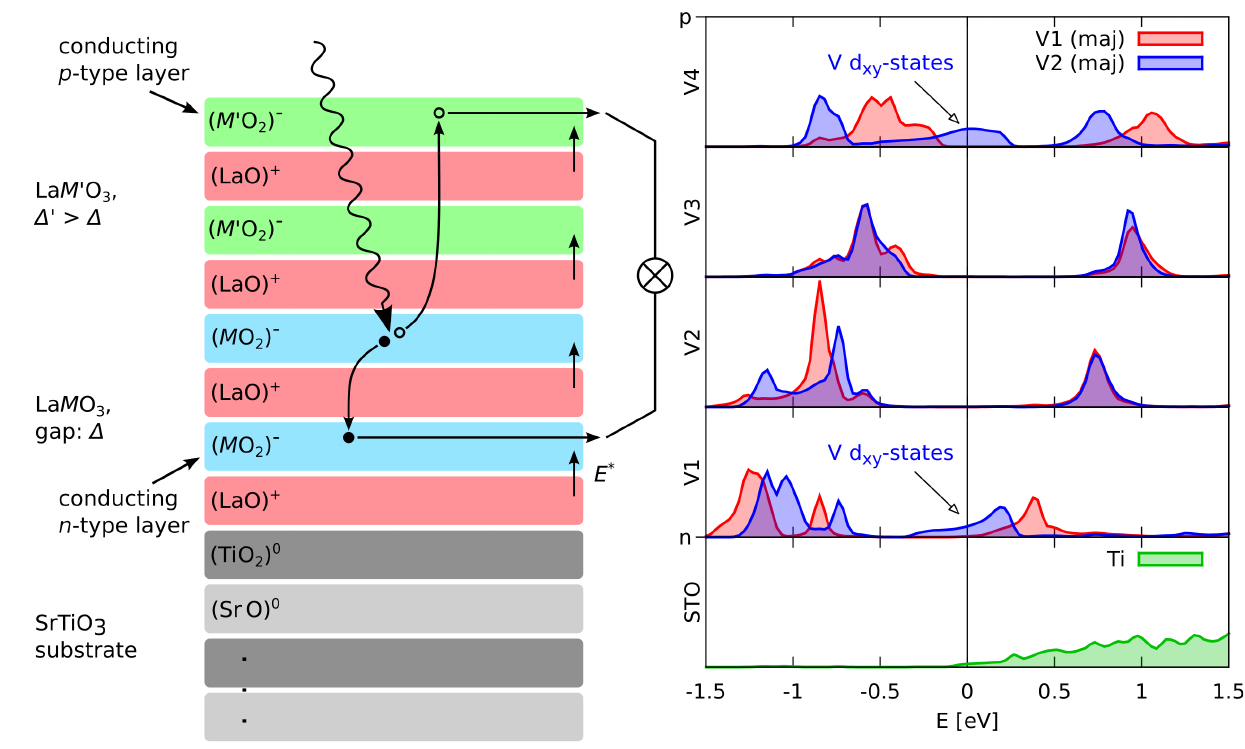}
\caption{Left: Proposal of solar cell made out of an oxide
heterostructure with a polar discontinuity. The emerging electric field
$E$ separates electrons and holes that are generated through the
photoelectric effect in the LaMO$_3$. Using different layers with e.g.
M=V, M'=Fe allows for a gap grading. Right: DFT+$U$ calculation for four
layer of LVO on top of a STO substrate (only the topmost STO layer is
shown).  The layer-resolved DOS shows the polar field as a shift of the
upper and lower Hubbard bands and an electronic reconstruction giving
rise to metallic interface and surface layers (adapted from
\cite{assmann13}).
\label{fig:solar1}}
\end{figure}

Fig. \ref{fig:solar1} (left panel) shows the scheme proposed in Ref.
\cite{assmann13} for a solar cell with transition-metal oxide
perovskites as absorbing materials.  The main idea of the LaVO$_3$-based
solar cells is that electrons and holes produced by the incoming photon
can be efficiently separated and harvested due to the polarity of the
heterostructure with its  intrinsic potential gradient (indicated as an
electric field $E$ in  the left panel of Fig. \ref{fig:solar1}).  The
DFT+$U$ results for 4 LVO layers on top of a STO substrate are  shown in
Fig. \ref{fig:solar1} (right panel).  The position of the lower and
upper LVO Hubbard band shifts from layer to layer, indicating the polar
electric field which can separate photoelectrically induced electrons
and holes.  Above a critical thickness of four layers,  the valence and
conduction $d$-bands of vanadium touch the Fermi level, which leads to
an electronic reconstruction and a metallic interface and surface layer.
The latter might need protection from a capping layer to suppress
disorder effects and achieve a sufficient electron mobility.  In a solar
cell, these metallic layers may be exploited to transport the electron
and holes to the power consumer.

The optical absorption which indicates the efficiency of the
photoelectric field in the LVO|STO heterostructure was calculated in
Ref. \cite{assmann13}.  The results compare quite favorably with the
most modern thin-film solar cell materials.  Particular appealing is
that oxide heterostructures can overcome the infamous Schockley-Queisser
limit of 38\% efficiencies in two ways: (i) As indicated in Fig.
\ref{fig:solar1}  (left panel) different transition metals M and M'  may
lead to different gaps (LaFeO$_3$ and LaVO$_3$ for instance were
considered in Ref. \cite{assmann13}). Such a ``band-grading'' is very
flexible in oxide heterostructures and can overcome loosing the excess
photon energy beyond the band gap to photons (heating up the solar cell
instead of gaining electric energy).  (ii) A genuine correlation effect,
impact ionization,  creates additional electron-hole pairs on the 10 fs
time scale \cite{werner14} if the photon  energy is larger than twice
the band gap.

Recently, Wang, {\it et al.}, \cite{wang15} reported the realization of
an actual solar cell, where LaVO$_3$ serves as the light absorber.
Further work on the quality of the heterostructure samples is needed to
enhance the efficiency of the solar cell which still suffers from a poor
mean free path  is only comparable to the dawn of Si solar cells.  Also
for the other proposed material, LaFeO$_3$ on SrTiO$_3$, Nakamura {\em
et al.} \cite{nakamura16} reported its photovoltaic applicability.

While it is unclear at present whether oxide heterostructures may become
a viable  alternative to Si solar cells, the proposal and experimental
confirmation already served another purpose: It confirms the existence
of a polar field. The latter has been controversially debated recently,
in particular in view of the fact that oxygen vacancies counteract the
polar field and also may induce a metallic interface layer, e.g. in
LAO/STO. Demonstrating a actually working solar cell proves at a polar
field survives in various oxide heterostructures.

\subsection{\label{Section:ruthenates} Ruthenates}  
Bulk SrRuO$_3$  (SRO) is a rare example of a ferromagnetic conducting
oxide~\cite{koster12}, and there is a long standing  experimental
experience in growing films of this  perovskite: thin single-domain
films of high quality can be grown with a pure SrO termination, which is
assisted by the volatility of RuO$_3$ and RuO$_4$ oxides~\cite{bell63}.
A typical substrate is STO with its cubic lattice constant
$a=3.905$\,\r{A} giving rise to to a moderate compressive strain
($\sim$0.45\,\%) in SRO~\cite{koster12}.

Resistivity measurements on ultrathin SRO layers show a pronounced
dependence on the layer thickness: while the bulk behavior is recovered
in thick slabs of $\gtrsim$15 SRO layers, fewer layers show a reduced
Curie temperature and a strongly enhanced resistivity~\cite{toyota05}.
Drastic changes occur upon reaching the critical thickness of four
monolayers, at which SRO turns into a non-magnetic
insulator~\cite{xia09}.  Experimental attempts to stabilize
ferromagnetism in single (or a few) SRO layers by capping or applying
compressive/tensile strain remain unsuccessful so far.

Since such  ultrathin ferromagnetic films are important for prospective
(e.g. spintronics) applications, understanding the nature of the
metal-insulator transition and the absence of ferromagnetism is
mandatory.  Quite remarkably, neither the electronic ground state, nor
an antiferromagnetism or otherwise non-ferromagnetic phase are captured
by DFT: it yields a ferromagnetic state for all slabs thicker than one
SRO monolayer and fails to reproduce the insulating
state~\cite{rondinelli08}.  Similarly, DFT+$U$ yields a metal even for
sizable $U$ values~\cite{rondinelli08}.  Later studies showed that an
antiferromagnetic insulator can be stabilized in DFT+$U$ for the
spurious RuO$_2$ termination~\cite{mahadevan09}, which is however in
contrast to the experiment~\cite{rijnders04}.  Moreover, DFT+$U$ yields
antiferromagnetism for layers of up to eight SRO monolayers, i.e.\
substantially overestimates the experimental critical thickness.

These apparent shortcomings of DFT and DFT+$U$ call for a more realistic
treatment of electronic correlations within DFT+DMFT.  The latter
accounts for the behavior of bulk SrRuO$_3$~\cite{jakobi11,kim15}, and
although DMFT is a cruder approximation in 2D than in 3D, we can still
expect it to capture the physics of thin SRO layers including its
magnetism.  The influence of the SRO layer thickness onto the electronic
and magnetic ground states was studied very recently in
Ref.~\cite{si15}.  All DFT calculations were done using a
$\sqrt{2}$$\times$$\sqrt{2}$$\times$6 supercell, which allows us to
study  SRO/STO compositions ranging from a SRO monolayer (1:5) to a
four-layer slab (4:2) and is compatible with both ferromagnetic and
antiferromagnetic in-plane order.  The out-of-plane unit cell parameter
as well as the internal atomic coordinates were optimized using DFT,
while keeping the in-plane lattice constant equal to that of the STO
substrate.   The wien2wannier\cite{kunes10}-derived  Hamiltonian  was
supplemented with a Coulomb repulsion $U$\,=\,3.5\,eV and Hund's
exchange $J$\,=\,0.3\,eV as derived by the constrained random phase
approximation (cRPA) approach.  The resulting interacting Hamiltonians
were treated in DMFT and solved using a CT-HYB solver.

\begin{figure}[t!]
\includegraphics[width=\columnwidth]{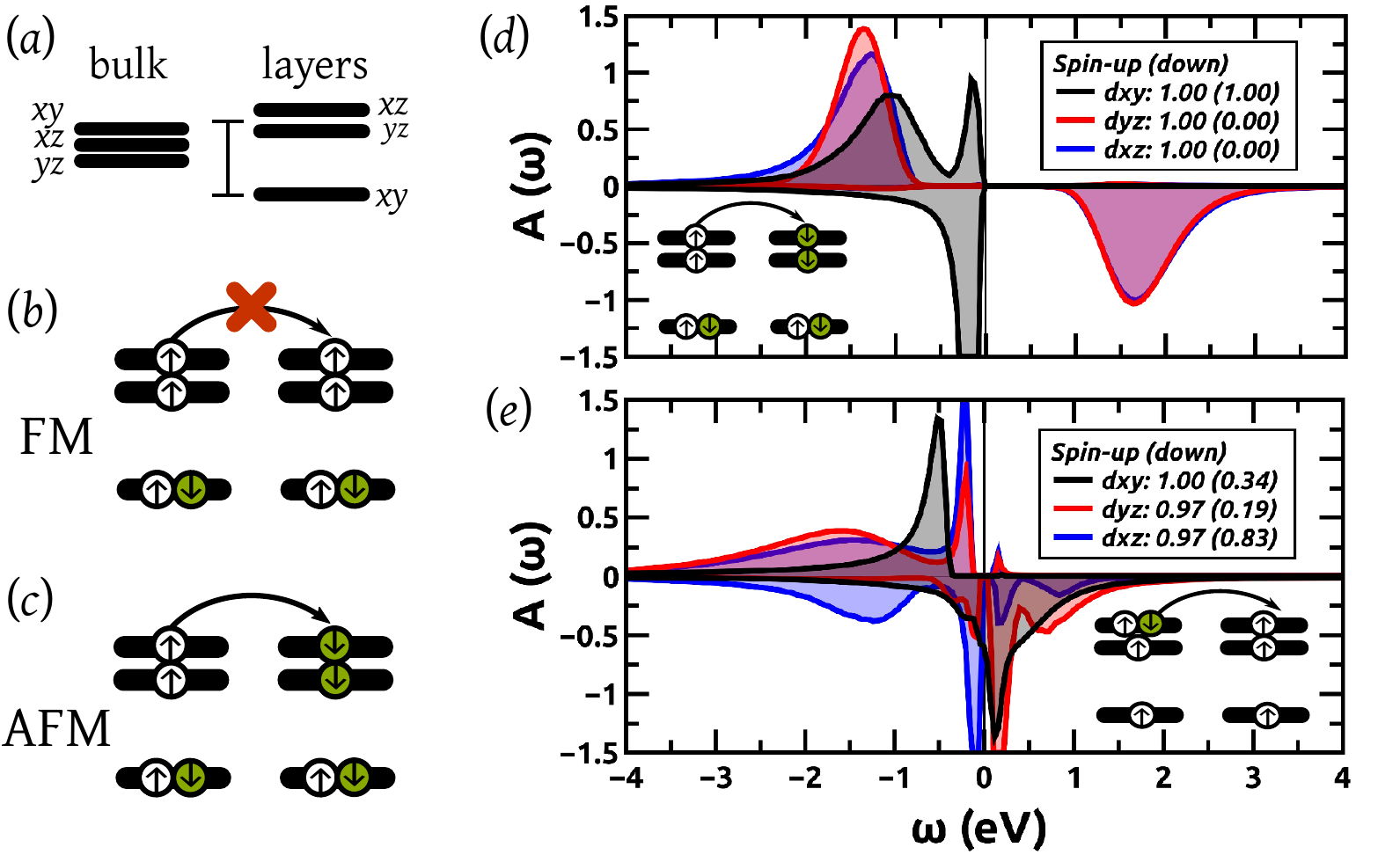}
\caption{
\label{fig:ru}
Ramifications of the dimensional reduction in SrRuO$_3$ thin films. (a)
The degeneracy of the three $t_{2g}$ in the bulk is lifted for ultrathin
layers.  (b) For a ferromagnetic configuration there is hence no
hopping, while for (c) an antiferromagnetic configuration virtual
(superexchange) hoppings within the $t_{2g}$ manifold are allowed. This
explains why antiferromagnetism is energetically favorable in SRO thin
films.  (d-e) Orbital-resolved DFT+DMFT spectral function $A$ as a
function of real frequency $\omega$ for (d) one SRO monolayer on a STO
substrate revealing an antiferromagnetic insulating state, and  (e) an
electron-doping with 4.3electrons/Ru turning the SRO film into a
half-metallic ferromagnet (adapted from
\cite{si15}).
}
\end{figure}

For SRO monolayers and bilayers, DFT+DMFT readily yields an insulating
antiferromagnetic state at room temperature (Fig.~\ref{fig:ru} d).  The
microscopic origin of this state lies in the sizable orbital
polarization: the $xy$ orbital is occupied by two electrons, while the
$xz$ and $yz$ orbitals are half-filled, see Fig.~\ref{fig:ru} a-c.  In
accord with the Hund's rule, the spins of the latter two orbitals align
parallel, giving rise to a localized magnetic moment of
$\sim$2\,$\mu_{\text{B}}$. While in a ferromagnetic configuration the
$xz$ and $yz$ electrons are essentially immobile (Fig.~\ref{fig:ru} b),
in the antiferromagnetic phase they can hop between the neighboring Ru
sites gaining kinetic energy through superexchange processes
(Fig.~\ref{fig:ru} c).  Hence DMFT does not only agree with  experiment,
but also explains the metal-insulator and
ferromagnetic-antiferromagnetic transition upon reducing the  thickness
of SRO films.

Unlike DFT and DFT+$U$, DMFT allows us to study the temperature
evolution of electronic and magnetic states.  In this way, it was shown
that SRO monolayers retain a sizable orbital polarization even in the
paramagnetic state ($\sim$1000\,K).  Indeed, a strong tendency towards
orbital polarization is visible already at the DFT level, marked by the
difference between the on-site energies of the $xy$ orbital and the
degenerate $xz$ and $yz$ orbitals.  This situation is typical for thin
films confined along the $z$ direction but apparently different from
bulk SRO, where the three $t_{2g}$ orbitals are degenerate
(Fig.~\ref{fig:ru} a).  This effect is somewhat countered by the
elongation of the topmost RuO$_6$ octahedra, which in the crystal-field
picture lowers the energy of $xz$ and $yz$ orbitals.  However, this
effect is weak and the $xy$ orbital always lies lower in
energy~\cite{si15}. 

Can DFT+DMFT calculations give a clue to the quest of growing ultrathin
{\em ferromagnetic} SRO films?  A standard way to manipulate physical
properties of thin films is the strain induced by the lattice mismatch
between the film and the substrate.  Unfortunately, DMFT calculations
reveal that for SRO neither compressive nor tensile strain stabilizes an
FM state, at least for realistic values of the on-site Coulomb
repulsion~\cite{si15}.  Another possibility to bring on-site energies
closer together is capping SRO with STO layers, which should to a
certain extent restore the bulk behavior.  Although the level splitting
is indeed reduced in DFT, DMFT still yields an antiferromagnetic
state~\cite{si15}, albeit with now somewhat more balanced orbital
occupations.  Thus, the antiferromagnetism is a robust feature of SRO
layers: neither strain nor capping can stabilize the desired
ferromagnetic state.  This explains why experimental attempts to
fabricate ferromagnetic SRO layers were hitherto unsuccessful.

There is however an alternative route to restore ferromagnetism: doping.
To verify this scenario, additional DFT calculations were performed
within the virtual crystal approximation, mimicking the Ru $d$-level
occupation of 3.7 electrons.  For this filling, DMFT indeed yields a
half-metallic ferromagnet (Fig.~\ref{fig:ru} e). Quite remarkably, for
both, electron and hole doping, the magnetic moments and the magnetic
ordering temperatures is higher than  in bulk SRO~\cite{si15}.

The good agreement between DFT+DMFT and experiment for the undoped phase
leaves little doubt that DMFT provides a realistic description of thin
SRO layers.  Still, there are many open issues that deserve further
studies with computational methods, including DMFT.  The first and
arguably most relevant is the effect of oxygen vacancies.  Real
materials are never perfect, and particularly thin films are often
subject to surface contaminations, growth defects and vacancies.  Oxygen
vacancies represent a severe experimental challenge: their concentration
is difficult to estimate, while the presence of a single vacancy can
dramatically alter the properties of the surrounding atoms.  Structural
relaxations invoked by vacancies can be studied by DFT, yet the
alteration of the Ru $4d$-level filling and its ramifications pertain to
correlated electrons and require an appropriate many-body treatment.
DFT+DMFT is a state-of-the-art tool for such problems.

\section{Conclusion and outlook}
\label{Section:conclusion}
DMFT is state-of-the-art for dealing with electronic correlations which
are of prime importance for many oxide heterostructures; and its merger
with DFT~\cite{kotliar06,held07} allows us to do realistic materials
calculations.  This way quantum correlations in time are included,
corresponding to a local self energy which might differ from site to
site. Physically, DFT+DMFT reliably describes among others,
quasiparticle-renormalizations, metal-insulator transitions, correlation
enhanced crystal field splittings, charge transfers, and magnetism. On
the other hand, non-local correlations are neglected. These are
important in the vicinity of phase transitions or to describe the
physics of excitons and weak localization corrections to the
conductivity. In the future, cluster~\cite{maier05} or diagrammatic
extensions~\cite{toschi07,rubtsov08} of DMFT, which are computationally
more demanding, and  hence hitherto focused on model Hamiltonians, will
also be applied to oxide heterostructures.

In the first part, Section \ref{Section:DFTDMFT}, we illustrated all
steps of a DFT+DMFT calculation for a showcase heterostructure: two
layers of SVO on an STO substrate.  This starts with a DFT calculation,
here using Wien2K~\cite{schwarz02}.  Subsequently a projection onto
Wannier orbitals and a corresponding low energy Hamiltonian is performed
with the help of wien2wannier~\cite{kunes10} (integrated in the most
recent Wien2K version) and Wannier90~\cite{mostofi08}. This may result
in a $d$-only Hamiltonian but also additional e.g. oxygen $p$ orbitals
can be included. The Wannier projection also defines what is
(site-)local in DMFT. For the subsequent DMFT calculation we used
w2dynamics~\cite{parragh12},\footnote{w2dynamics  will be made available
by Gnu Public License next year, betatest versions are available upon
email request.} which solves the DMFT impurity problem by
continuous-time quantum Monte Carlo simulations~\cite{gull11} in the
hybridization expansion.  This yields the DFT+DMFT self energy and
spectrum. For calculating  (optical) conductivities and thermal
responses, postprocessing with woptics\cite{assmann16} is needed;  a
DFT+DMFT charge self consistency is possible\cite{bhandary16}.

In the second part, Section \ref{Section:applications}, we reviewed
previous DFT+DMFT calculations for oxide heterostructures.  Not only
experimentally but also regarding DMFT calculations arguably most well
studied are titanates, in particular, LAO/STO and LTO/STO. Here the
former is only weakly correlated whereas electronic correlations are
strong for the latter as LTO is a Mott insulator. DMFT describes the
formation of a two dimensional electron gas at the interface due to an
electronic reconstruction, but also the localization of electrons at
oxygen vacancies if these are included in the supercell. 

Nickelate-based heterostructures may be used to turn the nickel Fermi
surface into a cuprate-like one~\cite{hansmann10}. This might however be
prevented by charge transfer processes: i.e. oxygen ligand holes may
lead to a $d^8L$ instead of a $d^7$ configuration~\cite{han11}. The
physics is completely different~\cite{parragh13}, spin-1 instead of
spin-1/2, and the theoretical uncertainty regarding the oxygen-$p$
position is too large to reliably predict which scenario prevails. The
smoking gun experiment in this respect is measuring the short-time
magnetic moment (spin) e.g. by neutron scattering or x-ray absorption
spectroscopy.

Turning to vanadates, LVO/STO is a promising candidate for efficient
solar cells because of the size of the LVO Mott-Hubbard gap and the
polar electric field of the heterostructure which has been
suggested~\cite{assmann13} and experimentally verified~\cite{wang15} to
separate photovoltaically generated electron-hole pairs. SVO on the
other hand is a strongly correlated metal in the bulk. With a
dimensional reduction it turns insulating below a critical thickness of
$\sim 3$ layers~\cite{yoshimatsu10,yoshimatsu11}.
DFT+DMFT~\cite{zhong15} could trace back the microscopic origin of this
metal-insulator transition to the correlation enhanced crystal field
splitting and showed that this transition can also be triggered by a
small gate voltage. This makes SVO/STO an ideal candidate for a Mott
transistor.

Ruthenates  such as SRO/STO are  promising for heterostructures that are
both metallic and ferromagnetic. DFT+DMFT~\cite{si15} could explain why
the experimental efforts to get  ultrathin ferromagnetic SRO/STO failed
so-far, despite predictions of the contrary: While bulk SRO is a
metallic ferromagnet in DMFT, the correlation enhanced crystal field
splitting makes  a competing antiferromagnetic phase favorable. This
antiferromagnetic phase is quite stable against strain, capping layers
etc. The most promising route to metallic ferromagnetism is electron
doping.

We have only seen the beginning of DFT+DMFT calculations for oxide
heterostructures. In the future, these calculation will help us to
better understand experiment and to predict novel physics, which
already led to some successes in the past. While the numerical effort is
larger than DFT, it is still much more efficient and cheaper than
experiment to scan the myriad of possible  combinations of
hetrerostructure slabs by DFT+DMFT. in experiment and necessitate a
refinement of the DFT+DMFT calculation.  In the past the theoretical and
experimental focus has been on heterostructures with layers
perpendicular to the (001) direction.  Now both experiment and theory
turn to other geometries:  confinement in the (110) direction  shows a
much more complicated quantization  and flat bands \cite{wang14};
bilayers in a (111) stacking on the other hand are interesting
regarding prospective topological states~\cite{xiao11}.  With the  close
cooperation between experiment on the one side and  DFT as well as DMFT
theory on the other side, a bright future in the research area of oxide
heterostructures lies ahead.

We thank  our colleagues and coauthors E. Assmann, S. Bhandary, P.
Blaha, P. Hansmann, R. Laskowski, G. Li, S. Okamoto, N. Parragh, L. Si,
J. Tomczak, A. Toschi, and M. Wallerberger for useful discussions and
joint efforts. This work was supported financially by European Research
Council under the European Union's Seventh Framework Program
(FP/2007-2013)/ERC grant agreement n.\ 306447,  the Austrian Science
Fund (FWF) through  SFB ViCoM F41 and project I-610, as well as research
group FOR1346 of the Deutsche Forschungsgemeinschaft (DFG). Calculations
reported were performed on the Vienna Scientific Cluster (VSC).

\end{document}